\documentclass[aps,prl,groupedaddress,fleqn,twocolumn]{revtex4}
\pdfoutput=1
\usepackage{graphicx}
\usepackage{amsmath}
\usepackage{gensymb}

\begin{document}

\title{Anomalous vacuum energy and stability of a quantum liquid}
\author{K. Trachenko$^{1}$}
\author{V. V. Brazhkin$^{2}$}
\address{$^1$ School of Physics and Astronomy, Queen Mary University of London, Mile End Road, London, E1 4NS, UK}
\address{$^2$ Institute for High Pressure Physics, RAS, 142190, Moscow, Russia}

\begin{abstract}
We show that the vacuum (zero-point) energy of a low-temperature quantum liquid is a variable property which changes with the state of the system, in notable contrast to the static vacuum energy in solids commonly considered. We further show that this energy is inherently anomalous: it decreases with temperature and gives negative contribution to system's heat capacity. This effect operates in an equilibrium and macroscopic system, in marked contrast to small or out-of-equilibrium configurations discussed previously. We find that the negative contribution is over-compensated by the positive term from the excitation of longitudinal fluctuations and demonstrate how the overall positive heat capacity is related to the stability of a condensed phase at the microscopic level.
%Instabilities and possibilities of entering new states in low-temperature liquids due to quantum fluctuations give rise to several prominent effects, stimulating the ongoing search for new types of quantum liquids \cite{annett,lee,balents,spin1,spin2,spin3,spin4,itou}. Instabilities are often related to negative heat capacity in non-equilibrium or small systems \cite{cv1,cv2,cv3,cv4}. Here, we show that the vacuum (zero-point) energy of any low-temperature quantum liquid is a dynamical property which is inherently anomalous: it decreases with temperature and gives negative contribution to system's heat capacity. This effect operates in an equilibrium and macroscopic system, in marked contrast to small or out-of-equilibrium configurations discussed previously \cite{cv1,cv2,cv3,cv4}. We show that the negative contribution is over-compensated by the positive term from the excitation of longitudinal fluctuations, and demonstrate how the overall positive heat capacity is related to the stability of a condensed phase at the microscopic level.
\end{abstract}

%\pacs{61.43.Fs, 64.70.Pf, 61.20.Lc}

\maketitle

The concept of zero-point (vacuum) energy arises in several fields of research and is related to notable fundamental effects as well as profound open problems, including in quantum mechanics, field theory and cosmology. Effects of zero-point fluctuations are also of interest in condensed matter physics and can be related to instabilities against forming new phases with interesting properties. For example, large energy of zero-point fluctuations de-stabilizes solidification of liquid He at room pressure, enabling the superfluidity to set in at low temperature \cite{annett}. Similarly, zero-point fluctuations are thought to prevent ordering in quantum spin liquids, potentially new quantum states of matter discussed more recently \cite{lee,balents,spin1,spin2,spin3,spin4,itou}.

The zero-point energy, $E_0$, is:

\begin{equation}
E_0=\sum\limits_{i=1}^n\frac{\hbar\omega_i}{2}
\label{defi}
\end{equation}

\noindent where $\omega_i$ is the mode frequency and $n$ is the number modes at operation.

In commonly discussed physical systems including in condensed matter and solid state theory, the vacuum energy is considered to be a constant, static, quantity. For example, in solids $n$ is fixed, resulting in constant $E_0$. For this reason, the vacuum energy does not affect measurable properties such as specific heat \cite{landau}.

Here, we show that the vacuum energy in quantum liquids is not a constant property. Instead, $E_0$ varies with the state of the system (temperature and pressure) due to the variation of the spectrum of transverse modes. This changes the general outlook on the vacuum energy.

More specifically, we show that $E_0$ {\it decreases} with temperature, resulting in the anomalous negative contribution to the specific heat. We note that negative specific heat often signals the onset of an instability, hence our finding represents another way in which zero-point fluctuations can have a de-stabilizing effect in addition to those mentioned earlier. We further show that the negative heat capacity term is over-compensated by the positive term due to longitudinal fluctuations, and demonstrate that the over-compensation is related to the stability of a condensed phase as such.

Interestingly, negative specific heat has been found to operate in small or out-of-equilibrium systems \cite{cv1,cv2,cv3,cv4,cv5}. In marked contrast to this, we show that the negative heat capacity component in a low-temperature quantum liquid is essentially a macroscopic and equilibrium effect related to the decrease of the vacuum energy of transverse fluctuations with temperature.

The development of general theory of liquids has been complicated by the combination of strong interactions and dynamic disorder, with the result that the liquid energy strongly depends on the type of interactions and therefore is system-specific, in contrast to solids and gases for which general expressions for the energy can be derived \cite{landau}. Perhaps surprisingly, considering low-temperature quantum liquids turns out to be simplifying feature for the theoretical description. As Landau argued, any weakly perturbed state of the quantum system is a set of elementary excitations, or quasi-particles. In the quantum liquid close to zero temperature, the quasi-particles are phonons, and are the lowest energy states in the system \cite{landau}. Consequently, theories of liquid helium and superfluidity involve phonons.

Notably, Landau considered longitudinal modes only because it was believed that a liquid does not support transverse modes. However, Frenkel earlier proposed that this is not the case: any liquid can also support transverse modes, albeit with frequencies above $\omega_{\rm F}$:

\begin{equation}
\omega>\omega_{\rm F}=\frac{1}{\tau}
\label{0}
\end{equation}

\noindent where $\tau$ is liquid relaxation time, the time between two consecutive particle jumps at one point in space \cite{frenkel} (see Figure 1).

\begin{figure}
\begin{center}
{\scalebox{0.6}{\includegraphics{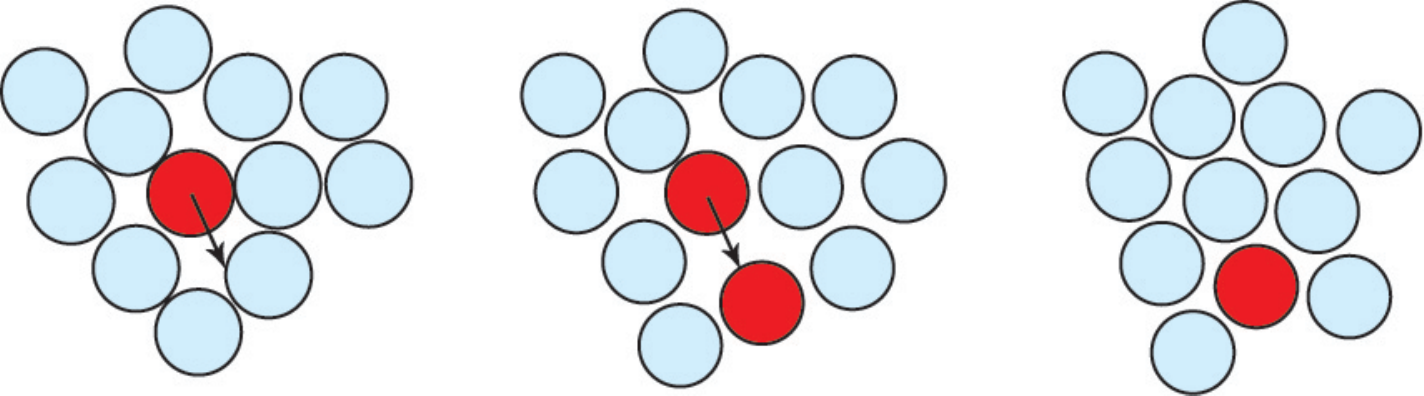}}}
\end{center}
\caption{Colour online. Illustration of a particle jump between two quasi-equilibrium positions in a liquid. These jumps take place with a period of $\tau$ on average.}
\label{jump}
\end{figure}

This prediction was based on a simple observation that at time shorter than $\tau$, the system does not flow, is a solid, and therefore supports all three modes including two transverse modes. Hence, a liquid supports two solid-like transverse modes with frequency above $\frac{1}{\tau}$. At frequency smaller than $\omega_{\rm F}$, transverse modes are non-propagating, and the liquid supports the hydrodynamic longitudinal mode only \cite{frenkel}.

It has taken long time to verify this prediction experimentally \cite{ropp,review}. Observed in viscous liquids considerably above the melting point (see, e.g., Refs. \cite{grim,scarponi}), the propagation of transverse modes was later studied in low-viscosity liquids on the basis of widely measured positive dispersion \cite{burkel,rec-review}. These studies included water \cite{water-fast}, where it was found that the onset of transverse excitations coincides with the inverse of liquid relaxation time \cite{water-tran}, as predicted by Frenkel. More recently, transverse modes were directly measured in the form of distinct dispersion branches and verified on the basis of computer modeling \cite{hoso,mon-na,mon-ga,sn,hoso3}. It is now well established that liquids sustain propagating solid-like modes, both longitudinal and transverse, extending to wavelengths comparable to interatomic separations.

We are therefore compelled to include transverse modes in the consideration of excitations in a quantum liquid on par with the longitudinal mode, and it is here where we find interesting and unexpected insights.

We consider the most general case of a low-temperature liquid where all modes are operative including the transverse ones. In a specific liquid system, the presence of transverse modes depends on temperature and pressure because $\tau$ in (\ref{0}) depends on both parameters \cite{ropp,condmat}. In commonly discussed liquid helium, transverse modes were not seen at ambient pressure but indirect evidence suggests their presence at elevated pressure \cite{jap}. Our main result (the variable vacuum energy) applies to any liquid although, as discussed below, it is most pronounced at low temperature where competing phonon excitations of transverse modes can be neglected.

As discussed earlier, liquid collective modes include one longitudinal mode and two transverse modes with frequency above $\omega_{\rm F}$ (the longitudinal mode is considered unmodified in this approach apart from its different propagation length in the hydrodynamic regime $\omega\tau<1$ and solid-like elastic regime $\omega\tau>1$ \cite{frenkel}). The mode (Planck) energy is $E_m=\frac{\hbar\omega}{2}+\frac{\hbar\omega}{\exp\frac{\hbar\omega}{T}-1}$ (here and below, $k_{\rm B}=1$). In the zero-temperature limit, the system's energy is the sum of the first terms of $E_m$, or $E_0$ in (\ref{defi}). At a finite temperature, the system's energy receives a contribution from the energy of mode excitations, $E_T$. $E_T$ is equal to the sum of the second terms of $E_m$ and explicitly depends on temperature. Then, the energy of the quantum liquid is

\begin{equation}
\begin{aligned}
&E=E_0+E_T\\
&E_0=E_0^l+E_0^t(\omega>\omega_{\rm F})\\
&E_T=E^l_T+E^t_T(\omega>\omega_{\rm F})
\end{aligned}
\label{1}
\end{equation}

\noindent

In (\ref{1}), $E_0$ is the vacuum (zero-point) energy of the liquid and includes the vacuum energy of the longitudinal mode, $E_0^l$, and the vacuum energy of two transverse modes with frequency above $\omega_{\rm F}$, $E^t_0(\omega>\omega_{\rm F})$. $E_T$ is the energy of mode excitations and includes the energy of excited longitudinal mode, $E^l_T$, and the energy of two excited transverse modes with $\omega>\omega_{\rm F}$, $E^t_T(\omega>\omega_{\rm F})$.

$E_0$ and $E_T$ can be calculated as the sums of the first and second term of $E_m$. Note that $E_T$ explicitly depend on temperature, whereas $E_0$ depends on temperature via $\omega_{\rm F}$ as discussed below.

We start with the vacuum energy in (\ref{defi}) and (\ref{1}), $E_0=E_0^l+E_0^t(\omega>\omega_{\rm F})$. The two terms can be calculated as $E_0^l=\int\limits_0^{\omega_{\rm D}}\frac{\hbar\omega}{2}g_l(\omega)d\omega$ and $E_0^t=\int\limits_{\omega_{\rm F}}^{\omega_{\rm D}}\frac{\hbar\omega}{2}g_t(\omega)d\omega$, where $N$ is the number of particles, $g_l(\omega)=\frac{3N}{\omega_{\rm D}^3}\omega^2$ and $g_t(\omega)=\frac{6N}{\omega_{\rm D}^3}\omega^2$ are longitudinal and transverse phonon density of states, respectively, and where we assumed, without the loss of generality, that the maximal frequency of longitudinal and transverse modes is approximately the same and given by Debye frequency $\omega_{\rm D}$. The lower integration limit in $E_0^t$, ${\omega_{\rm F}}$, signifies that the spectrum of transverse modes starts with ${\omega_{\rm F}}$ as discussed above ($\omega_{\rm F}<\omega_{\rm D}$). Integrating gives the vacuum energy as

\begin{equation}
E_0=\frac{3}{8}N\hbar\omega_{\rm D}+\frac{3}{4}N\hbar\omega_{\rm D}\left(1-\left(\frac{\omega_{\rm F}}{\omega_{\rm D}}\right)^4\right)
\label{2}
\end{equation}

$\omega_{\rm F}=\frac{1}{\tau}$ in (\ref{2}) depends on pressure and temperature: relaxation time $\tau$ decreases with temperature and increases with pressure. Therefore, the vacuum energy of liquids is a variable property which changes with the state of the system.

According to (\ref{2}), the vacuum energy $E_0$ decreases with temperature because $\omega_{\rm F}=\frac{1}{\tau}$ increases ($\tau$ decreases with temperature). This reflects the decrease of the number of transverse modes because they propagate only above frequency $\omega_{\rm F}=\frac{1}{\tau}$ (see Eq. (\ref{0})). From (\ref{2}), the constant-volume specific heat $c_0=\frac{1}{N}\frac{dE_0}{dT}$ due to the vacuum energy is

\begin{equation}
c_0=-3\hbar\left(\frac{\omega_{\rm F}}{\omega_{\rm D}}\right)^3\frac{d\omega_{\rm F}}{dT}
\label{3}
\end{equation}

We observe that $c_0$ in (\ref{3}) is negative because $\frac{d\omega_{\rm F}}{dT}>0$ as noted above.

We make two observations regarding the unusual negative sign of $c_0$. First, this is a purely quantum effect, and is absent in the classical case. Indeed, the mode energy is $E_m=T$ in the classical case. Consequently, the decrease of the number of two transverse modes with temperature according to (\ref{0}) leads to their energy changing from $2NT$ at low temperature to $NT$ at high, corresponding to the potential energy of transverse modes becoming zero \cite{prb,ropp,review}. Adding the energy of the longitudinal mode, $NT$, we find that the liquid specific heat changes from 3 to 2, in agreement with the experimental results \cite{prb,ropp,review}. Hence, the energy of the classical liquid always increases with temperature, although the energy slope and specific heat decreases from 3 to 2. In contrast, the vacuum energy of the quantum liquid does not contain the $\propto T$ term as in the classical case, and decreases with temperature in (\ref{2}) as a result.

Second, previous discussions have attributed negative heat capacity to the non-equilibrium state of the system or its smallness in terms of the number of atoms \cite{cv1,cv2,cv3,cv4,cv5}. In marked contrast to this, negative $c_0$ due to the vacuum energy operates in a macroscopic and equilibrium quantum liquid: the decrease of the vacuum energy operates in an equilibrium gas of phonon excitations in an arbitrarily large system.

We now show that the total specific heat of the quantum liquid is necessarily positive. This follows from the calculation of the remaining energy term related to phonon excitations in (\ref{1}), $E_T=E^l_T+E^t_T(\omega>\omega_{\rm F})$. We consider the low-temperature regime $T\ll\hbar\omega_{\rm D}$. $E^l_T$ at low temperature can be calculated as $E^l_T=\int\limits_0^{\omega_{\rm D}}\frac{\hbar\omega}{\exp\left(\frac{\hbar\omega}{T}\right)-1}g_l(\omega)d\omega=NTD\left(\frac{\hbar{\omega_{\rm D}}}{T}\right)$, where $D(x)=\frac{3}{x^3}\int\limits_0^x\frac{z^3{\rm d}z}{\exp(z)-1}$ is Debye function. Taking the low-temperature limit $T\ll\hbar\omega_{\rm D}$ where $D(x)=\frac{\pi^4}{5x^3}$ \cite{landau} gives $E^l_T=\frac{N\pi^4T^4}{5\left(\hbar\omega_{\rm D}\right)^3}$. The same result follows from the exact calculation not relying on the Debye model and from noting that at low temperature, the upper integration limit can be extended to infinity due to the fast convergence of the integral \cite{landau}. Then, $c_v^l$ due to the longitudinal phonon excitations is

\begin{equation}
c_v^l=\frac{4\pi^4}{5}\left(\frac{T}{\hbar\omega_{\rm D}}\right)^3
\label{4}
\end{equation}

The last term in (\ref{1}), $E^t_T(\omega>\omega_{\rm F})$, representing the excitations of two transverse modes, is $E^t_T(\omega>\omega_{\rm F})=\int\limits_{\omega_{\rm F}}^{\omega_{\rm D}}\frac{\hbar\omega}{\exp\frac{\hbar\omega}{T}-1}g_t(\omega)d\omega$, and turns out to be zero at low temperature. This can be seen by writing $E^t_T(\omega>\omega_{\rm F})$ as

\begin{equation}
E^t_T(\omega>\omega_{\rm F})=\int\limits_0^{\omega_{\rm D}}\frac{\hbar\omega g_t(\omega)d\omega}{\exp\frac{\hbar\omega}{T}-1}-\int\limits_0^{\omega_{\rm F}}\frac{\hbar\omega g_t(\omega)d\omega}{\exp\frac{\hbar\omega}{T}-1}
\label{5}
\end{equation}

Integrating (\ref{5}) gives $E^t_T(\omega>\omega_{\rm F})=2NTD\left(\frac{\hbar\omega_{\rm D}}{T}\right)-2NT\left(\frac{\omega_{\rm F}}{\omega_{\rm D}}\right)^3D\left(\frac{\hbar\omega_{\rm F}}{T}\right)$. In the low-temperature limit $T\ll\hbar\omega_{\rm D}$ where $D(x)=\frac{\pi^4}{5x^3}$, the two terms cancel exactly, giving $E^t_T(\omega>\omega_{\rm F})=0$. The same result follows without relying on the Debye model and from observing that in the low-temperature limit, the upper integration limits in both terms in (\ref{5}) can be extended to infinity due to fast convergence of integrals. Then, $E^t_T(\omega>\omega_{\rm F})$ in (\ref{5}) is the difference between two identical terms and is zero.

Physically, the reason for $E^t_T(\omega>\omega_{\rm F})=0$ is that only high-frequency transverse modes exist in a liquid according to (\ref{0}), but these high-energy modes are not excited at low temperature.

Therefore, the total constant-volume specific heat of the low-temperature quantum liquid, $c_v$, is $c_v=c_0+c_v^l$, the sum of (\ref{3}) and (\ref{4}):

\begin{equation}
c_v=\frac{4\pi^4}{5}\left(\frac{T}{\hbar\omega_{\rm D}}\right)^3-3\hbar\left(\frac{\omega_{\rm F}}{\omega_{\rm D}}\right)^3\frac{d\omega_{\rm F}}{dT}
\label{6}
\end{equation}

We consider two common model temperature dependencies of $\tau$ ($\omega_{\rm F}$). The first is an exponential dependence $\tau=\tau_{\rm D}\exp\left(\frac{U}{T}\right)$, where $U$ is the activation energy barrier for an atomic jump shown in Figure 1 \cite{dyre} and $\tau_{\rm D}$ is Debye vibration period. The exponential dependence is seen in low-viscosity liquids \cite{dyre} and low-temperature Bose liquids including the normal flow component of liquid $^4$He \cite{donnelly}. The second dependence, the inverse power law, $\tau\propto\frac{1}{T^\alpha}$, is discussed \cite{pines} and observed \cite{he3-1,he3-2,he3-3} in quantum liquids with Fermi statistics, with typically $\alpha=2$ (below we consider $\alpha\ge 1$). Both models of $\tau$ operate in the range $T\ll U$ where the concept of $\tau$ applies and where particle dynamics can be separated into oscillations around quasi-equilibrium positions and jumps between these positions with period $\tau$ \cite{prl,phystoday}. The two models may not describe the experimental behavior of $\tau$ in a wide temperature range, however they are convenient for qualitative estimations.

For the exponential dependence, $\omega_{\rm F}=\omega_{\rm D}\exp\left(-\frac{U}{T}\right)$, and the total specific heat in (\ref{6}) becomes:

\begin{equation}
c_v=\frac{4\pi^4}{5}\left(\frac{T}{\hbar\omega_{\rm D}}\right)^3-3\hbar\omega_{\rm D}\exp\left(-\frac{4U}{T}\right)\frac{U}{T^2}
\label{61}
\end{equation}

At first glance, $c_v$ of the low-temperature liquid in (\ref{61}) can attain both positive and negative values. In fact, $c_v$ in (\ref{6}) is always positive, and there is a fundamental reason for it as discussed below.

To show that $c_v$ in (\ref{61}) is always positive, we need to demonstrate that the ratio of the absolute values of the first and the second term, $f$, is always larger than 1. From (\ref{61}), $f$ is

\begin{equation}
f=\frac{4\pi^4}{15}\frac{1}{U\left(\hbar\omega_{\rm D}\right)^4}T^5\exp\left(\frac{4U}{T}\right)
\label{7}
\end{equation}

$T^5\exp\left(\frac{4U}{T}\right)$ in (\ref{7}) increases to infinity at both small and large $T$, and has a minimum at $T_{\rm min}=\frac{4U}{5}$. Then, the minimal value of $f$, $f_{\rm min}$, is

\begin{equation}
f_{\rm min}=C\left(\frac{U}{\hbar\omega_{\rm D}}\right)^4
\label{8}
\end{equation}

\noindent where $C=\frac{2^{12}\pi^4e^5}{3\cdot 5^6}\approx 1264$.

We therefore find that $f_{\rm min}>1$ and hence $f>1$ always hold and ensure $c_v>0$, provided $\frac{U}{\hbar\omega_{\rm D}}>1$.

\begin{figure}
\begin{center}
{\scalebox{0.35}{\includegraphics{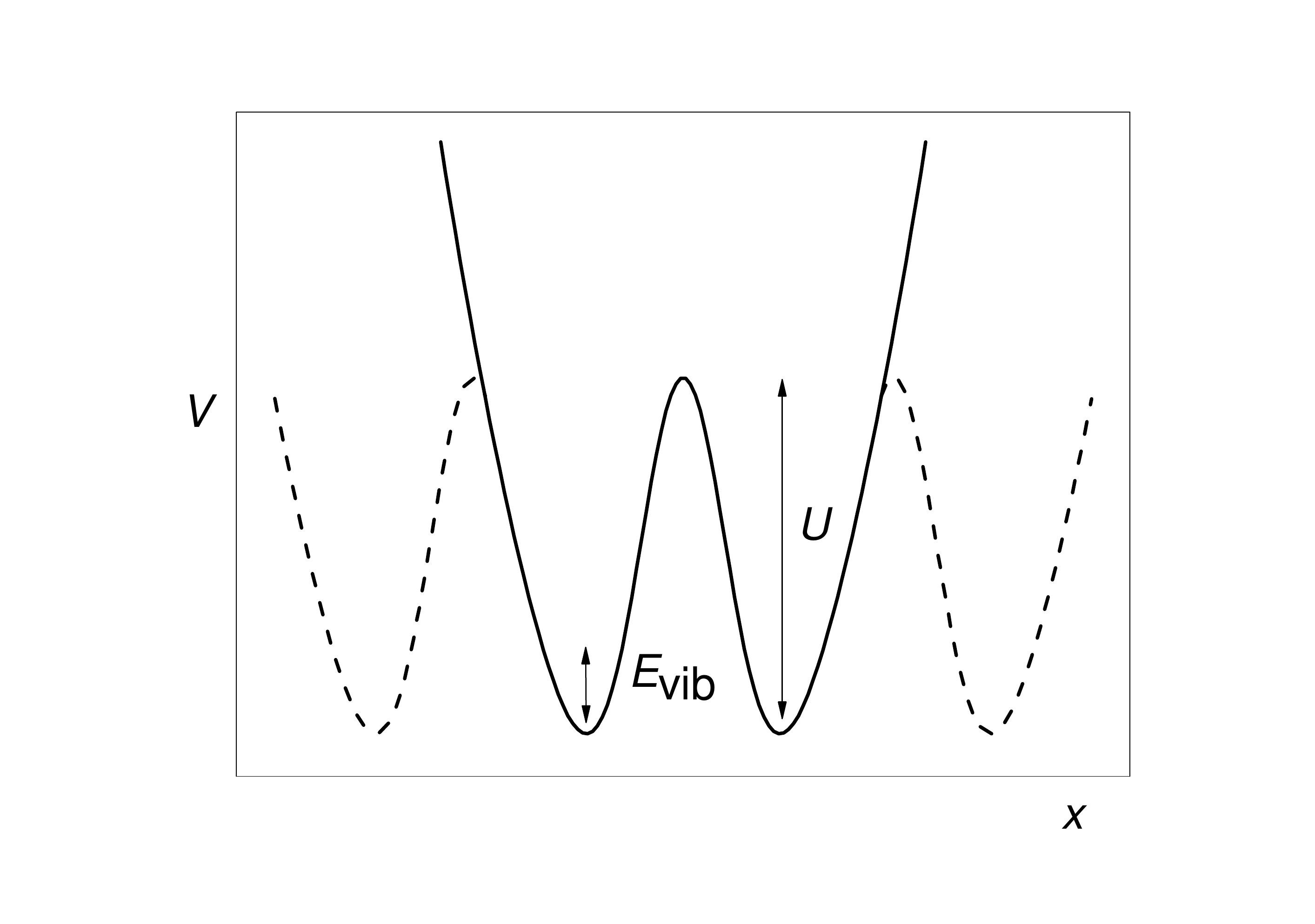}}}
\end{center}
\caption{Interaction potential $V$ in a condensed phase (liquid or solid) along a reaction coordinate $x$, illustrating that the activation energy $U$ is larger than the energy of vibrations $E_{\rm vib}\approx\hbar\omega_{\rm D}$. Dashed line implies that the potential is multi-well in the liquid leading to multiple diffusive jumps. Schematic illustration.}
\label{poten}
\end{figure}

The inequality $\frac{U}{\hbar\omega_{\rm D}}>1$ is an inherent feature of a stable particle configuration (in fact, $\frac{U}{\hbar\omega_{\rm D}}\gg 1$ as discussed below). Indeed, a cohesive state of condensed matter is related to the existence of a minimum of interatomic interaction energy. In a stable condensed state, particles oscillate around the potential minima. The energy of this motion is governed by the energy of interatomic interactions at small displacements, and is of the order of $\hbar\omega_{\rm D}$. On the other hand, the energy needed to break the particle cage, $U$, is governed by the energy of interatomic interactions but at much larger displacements comparable to interatomic separations (see Figure \ref{poten}). Therefore, the inequality $U\gg\hbar\omega_{\rm D}$ necessarily holds for any smooth interatomic potential increasing away from the minimum.

Therefore, the presence of the potential minima and related stability of any cohesive configuration implies $U>\hbar\omega_{\rm D}$ and ultimately leads to $c_v>0$. Reversing the argument, $c_v<0$ implies that $U>\hbar\omega_{\rm D}$ does not hold, in which case the system becomes unstable as a condensed phase and rearranges into a non-cohesive state such as, for example, a gas phase with different properties.

Another insight into the origin of inequality $U>\hbar\omega_{\rm D}$ follows form the consideration of characteristic energy scales in a stable condensed phase. A condensed phase exists due to the stability of electronic configurations between atoms, and any deformation of electronic density needed to break interatomic ``bonds'' to cause cage rearrangement involves the energy $U$ on the order of characteristic electronic energy $E_{\rm el}$: $U\approx E_{\rm el}$. Hence, $\frac{U}{\hbar\omega_{\rm D}}\approx\frac{E_{\rm el}}{\hbar\omega_{\rm D}}$. The last ratio, $\frac{E_{\rm el}}{\hbar\omega_{\rm D}}$, is known to be approximately equal to $\left(\frac{M_{\rm ion}}{m_{\rm el}}\right)^{\frac{1}{2}}$, where $M_{\rm ion}$ and $m_{\rm el}$ are ionic and electronic masses. Hence, $\frac{U}{\hbar\omega_{\rm D}}\gg 1$ (in numerical terms, $E_{\rm el}$ is typically on the order of eV $\approx 10^4$ K whereas $\hbar\omega_{\rm D}$ is in the range $100-1000$ K).

We now consider the inverse power law $\tau\propto\frac{1}{T^{\alpha}}$, which we write as
$\tau=\tau_{\rm D}\left(\frac{U}{T}\right)^{\alpha}$. This reflects the property that $\tau$ approaches its smallest value $\tau_{\rm D}$ when $T$ and $U$ become comparable and when particles spend approximately the same time in intra- and inter-valley motions in Figure \ref{poten} \cite{prl}. Similarly to the exponential dependence considered above, the power law applies in the range $T\ll U$. Then, writing $\omega_{\rm F}=\omega_{\rm D}\left(\frac{T}{U}\right)^{\alpha}$ and using it in (\ref{6}) gives

\begin{equation}
c_v=\frac{4\pi^4}{5}\left(\frac{T}{\hbar\omega_{\rm D}}\right)^3-\frac{3\alpha\hbar\omega_{\rm D}}{T}\left(\frac{T}{U}\right)^{4\alpha}
\label{9}
\end{equation}

As in the case of (\ref{61}), $c_v>0$ is ensured by $f>1$, where from (\ref{9}) $f$ is

\begin{equation}
f=\frac{4\pi^4}{15\alpha}\left(\frac{U}{\hbar\omega_{\rm D}}\right)^4\left(\frac{U}{T}\right)^{4\alpha-4}
\label{10}
\end{equation}

We observe that $\frac{U}{\hbar\omega_{\rm D}}$, the same factor as in (\ref{8}), features in (\ref{10}), and is always larger than 1 as discussed above. Then, with $T\ll U$ and $\alpha\ge 1$ as discussed above, $f$ in (\ref{10}) is larger than $1$ for any reasonable $\alpha$, ensuring $c_v>0$.

We therefore find that transverse modes give negative contribution to heat capacity of the quantum liquid, but the longitudinal excitation prevents the net heat capacity from becoming negative. Interestingly, no transverse modes can exist without the longitudinal mode (the opposite is not the case: there are systems with the longitudinal mode only but no transverse modes such as one-dimensional systems). Hence the longitudinal mode always ``protects'' the system's heat capacity from becoming negative.

We note that over-compensation of the negative term in $c_v$ and the overall positivity of $c_v$ take place even in the absence of additional contribution from exchange effects that can set in the liquid at low temperature \cite{ceperley}.

As mentioned above, our analysis of positivity of $c_v$ concerns the low-temperature regime $T\ll\hbar\omega_{\rm D}$ where transverse modes do not contribute to the liquid energy (see Eq. \ref{5} and accompanying discussion). At higher temperature, one needs to consider the increasing contribution of excited transverse modes and its competition with the decrease of the number of transverse modes according to (\ref{0}) \cite{ropp,review}. One can consider the high-temperature classical case where the decrease of $c_v$ due to progressively disappearing transverse modes is most pronounced because all phonons are excited. In this case, $c_v$ is positive (larger than 2 as discussed above) as in the low-temperature case due to $\frac{d\tau}{dT}$ tending to zero at high temperature \cite{prb}.

There is another fundamental reason why the anomalous decrease of the vacuum energy operates at low temperature only. As discussed above, this anomalous decrease is related to the reduction of the energy of transverse modes with temperature. This process operates in the regime where diffusive particle jumps between quasi-equilibrium points can be defined. In this regime, $\omega_{\rm F}<\omega_{\rm D}$, and the concept of $\tau$ ($\omega_{\rm F}$) and Eq. \ref{2} apply. When $\omega_{\rm F}$ becomes comparable to $\omega_{\rm D}$ at higher temperature, the liquid crosses the Frenkel line (FL) \cite{ropp} where the dynamical regime changes from combined oscillatory and diffusive to purely diffusive as in a gas. At this point, the evolution of collective modes in the system changes qualitatively because two transverse modes disappear completely  \cite{ropp} (on further temperature increase above the FL, the remaining longitudinal mode disappears starting from the shortest wavelength). In the gas-like regime above the FL, $T>U$ applies. Together with $U>\hbar\omega_{\rm D}$ as discussed above, this implies $T>\hbar\omega_{\rm D}$. This is the condition for the classical regime where the anomalous behavior of the vacuum energy is absent as discussed earlier.

Before concluding, we note that the negative heat capacity component discussed here has temperature dependence that is different from the longitudinal term. Therefore, this component could be detectable experimentally in principle. Although its relative weight is small in the overall heat capacity of liquids, new types of systems currently explored and expanding experimental conditions \cite{annett,lee,balents,spin1,spin2,spin3,spin4,itou} may bring interesting results.

In summary, the vacuum energy of liquids emerges as a variable property which changes with the state of the system, in notable contrast to the static vacuum energy in solids commonly considered. Moreover, this energy anomalously decreases with temperature, contributing negatively to system's heat capacity. Interestingly different from the previously known examples, this effect operates in an equilibrium and macroscopic system. The net heat capacity of quantum liquids is necessarily positive, and is intimately related to the stability of a condensed phase itself.

%K. T. is grateful to EPSRC and V. V. Brazhkin to RSF 14-22-00093 for support.

\end{document}